\documentclass[journal]{./IEEEtran}

\usepackage{cite}
\usepackage{booktabs}

\usepackage{lineno}
\usepackage{xspace}
\newcommand{\ItwoC} {I\textsuperscript{2}C\xspace}

\usepackage{hyperref}

\ifCLASSINFOpdf
   \usepackage[pdftex]{graphicx}
\else
   \usepackage[dvips]{graphicx}
\fi

\begin{document}

\title{FELIX: the New Detector Interface for \\the ATLAS Experiment}
%
%
% author names and IEEE memberships
% note positions of commas and nonbreaking spaces ( ~ ) LaTeX will not break
% a structure at a ~ so this keeps an author's name from being broken across
% two lines.
% use \thanks{} to gain access to the first footnote area
% a separate \thanks must be used for each paragraph as LaTeX2e's \thanks
% was not built to handle multiple paragraphs
%
%\author{Weihao~Wu,~\IEEEmembership{Member,~IEEE,}
%	John~Doe,~\IEEEmembership{Fellow,~OSA,}
%	and~Jane~Doe,~\IEEEmembership{Life~Fellow,~IEEE}% <-this % stops a space

%\author{Weihao Wu, Anamika Aggarwal, Kevin Bauer, Mikael Baymani, Andrea  Borga, Henk Boterenbrood, Hucheng Chen, Kai Chen, Mark Donszelmann, Frank Filthaut, Israel Grayzman, Daniel Guest,  Rene Habraken, Markus Joos, Serguei Kolos, Andy Lankford, Francesco Lanni, Giovanna Lehmann Miotto, Lorne Levinson, Julia Narevicius, Michael Oberling, Will Panduro, Alex Paramonov, Frans Schreuder, Jorn Schumacher, Alex Roich, Shaochun Tang, Gokhan Unel, Wainer Vandelli, Jos Vermeulen, Jinlong Zhang

\author{
	
%	A.\,Aggarwal, K.\,Bauer, M.\,Baymani, A.\,Borga, H.\,Boterenbrood, H.\,Chen, K.\,Chen, M.\,Donszelmann, F.\,Filthaut, I.\,Grayzman, D.\,Guest,  R.\,Habraken, M.\,Joos, S.\,Kolos, A.\,Lankford, F.\,Lanni, G.\,Lehmann Miotto, L.\,Levinson, J.\,Narevicius, M.\,Oberling, J.G.\,Panduro\,Vazquez, A.\,Paramonov, F.\,Schreuder, J.\,Schumacher, A.\,Roich, S.\,Tang, G.\,Unel, W.\,Vandelli, J.\,Vermeulen, W.\,Wu, J.\,Zhang
W.\,Wu \\on behalf of the ATLAS TDAQ Collaboration

%\thanks{Weihao Wu was with the Department
%	of Electrical and Computer Engineering, Georgia Institute of Technology, Atlanta,
%	GA, 30332 USA e-mail: (see http://www.michaelshell.org/contact.html).}% <-this % stops a space
%\thanks{Weihao Wu, Hucheng Chen, Kai Chen, Francesco Lanni and Shaochun Tang are with Brookhaven National Laboratory, P.O. Box 5000, Upton, NY 11973-5000, USA. }
%\thanks{Michael Oberling, Alex Paramonov and Jinlong Zhang are  with  Argonne  National  Laboratory,  Lemont,  IL  60439, USA. }
%\thanks{Andrea  Borga, Henk Boterenbrood, Frans Schreuder and Jos Vermeulen are  with  Nikhef  National  Institute  for  Subatomic  Physics  I University of Amsterdam. }
%\thanks{Anamika Aggarwal, Mark Donszelmann, Frank Filthaut and Rene Habraken are with Radboud University Nijmegen, Comeniuslaan 4, 6525 HP Nijmegen, Netherlands.}
%\thanks{Mikael Baymani, Markus Joos,  Giovanna Lehmann Miotto, Jorn Schumacher and Wainer Vandelli are with CERN, CH-1211 Geneva 23, Switzerland.}
%\thanks{Will Panduro is with University of London, Royal Holloway, UK.}
%\thanks{Kevin Bauer, Daniel Guest, Serguei Kolos, Andy Lankford and Gokhan Unel are with the University of California Irvine.}
%\thanks{Israel Grayzman, Lorne Levinson, Julia Narevicius and Alex Roich are with the Department of Particle Physics, The Weizmann Institute of Science, Rehovot 76100, Israel.}}

\thanks{W.\,Wu is with Brookhaven National Laboratory, P.O.\,Box 5000, Upton, NY 11973-5000, USA.\, Email: weihaowu@bnl.gov.}
%\thanks{Copyright 2018 CERN for the benefit of the ATLAS Collaboration. Reproduction of this article or parts of it is allowed as specified in the CC-BY-4.0 license.}}

\thanks{From ATL-DAQ-PROC-2018-003. Published with permission by CERN.}}

\maketitle

% As a general rule, do not put math, special symbols or citations
% in the abstract or keywords.
\begin{abstract}
During the next major shutdown (2019-2020), the ATLAS experiment at the LHC will adopt the Front-End Link eXchange (FELIX) system as the interface between the data acquisition, detector control and TTC (Timing, Trigger and Control) systems and new or updated trigger and detector front-end electronics. FELIX will function as a router between custom serial links from front-end ASICs and FPGAs to data collection and processing components via a commodity switched network. Links may aggregate many slower links or be a single high bandwidth link. FELIX will also forward the LHC bunch-crossing clock, fixed latency trigger accepts and resets received from the TTC system to front-end electronics. The FELIX system uses commodity server technology in combination with FPGA-based PCIe I/O cards. The FELIX servers will run a software routing platform serving data to network clients. Commodity servers connected to FELIX systems via the same network will run the new Software Readout Driver (SW ROD) infrastructure for event fragment building and buffering, with support for detector or trigger specific data processing, and will serve the data upon request to the ATLAS High Level Trigger for Event Building and Selection. This paper will cover the design and status of FELIX, the SW ROD, results of early performance testing and integration tests with several ATLAS front-ends.
\end{abstract}

% Note that keywords are not normally used for peerreview papers.

% For peer review papers, you can put extra information on the cover
% page as needed:
% \ifCLASSOPTIONpeerreview
% \begin{center} \bfseries EDICS Category: 3-BBND \end{center}
% \fi
%
% For peerreview papers, this IEEEtran command inserts a page break and
% creates the second title. It will be ignored for other modes.
\IEEEpeerreviewmaketitle

\section{Introduction}

% \IEEEPARstart{T}{his} demo file is intended to serve as a ``starter file''
% for IEEE journal papers produced under \LaTeX\ using
% IEEEtran.cls version 1.8b and later.

\begin{figure}[!t]
 	\centering
 	\includegraphics[width=3in]{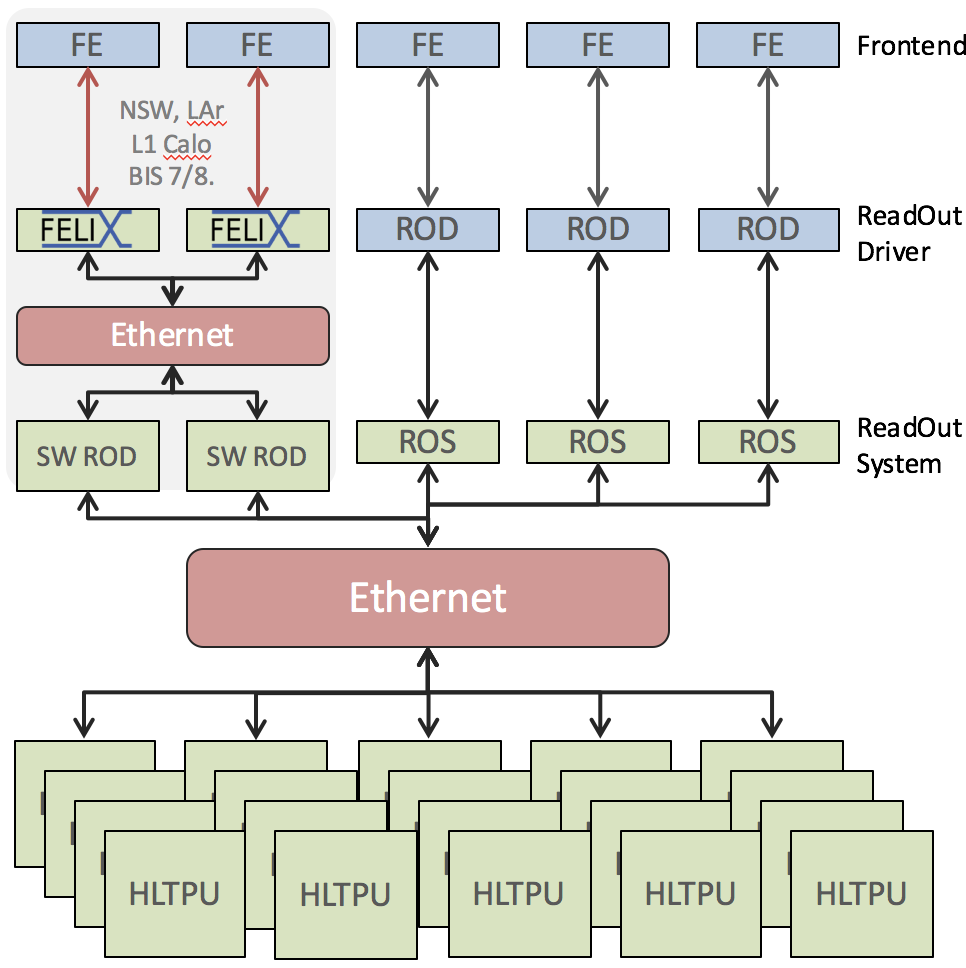}
 	\caption{ATLAS DAQ architecture in Run 3 upgrade}
 	\label{fig: felix_atlas}
 \end{figure}

%  It has a two stage trigger system in Run-2: the hardware-based leven-1 (L1) trigger system and software-based high level trigger (HLT) system. The L1 trigger system reduces the 40\,MHz collision rate to approximately 100 kHz within 2.5 microseconds. This L1 trigger is distrubted to on-deteector Front Ends electronics (FE) in fixed latency. The FE modules will forward accepted data streams to the back-end electronics (ReadOut Drivers, RODs) by means of custom point-to-point detector-specfic links.  The RODs perform data manipulation tasks like aggregation or compression before pushing the data to the circa 100 ReadOut System (ROS) PCs over 1800 point-to-point optical links (S-Link) [2]. ROS PCs buffer event fragments and forward them upon request to the High-Level Trigger (HLT) compute farm, consisting of 1500 servers. Here, the event rate is further reduced to the target event rate of about 1 kHz. In this two-stage trigger and DAQ system, the limitations identified included the difficulty to overcome data congestion in the RODs or ROBs as well as lack of scalability.
 \IEEEPARstart{T}{he} Large Hadron Collider (LHC) will undergo a series of significant upgrades in the next ten years, which increase both collision energy and peak luminosity. As one of the four major experiments, the ATLAS experiment will also follow the same upgrade steps\cite{lhc}. The Front End LInk eXchange (FELIX) is a new detector readout component being developed as part of the ATLAS upgrade effort\,\cite{felix_andrea}. FELIX is designed to act as a data router, receiving packets from detector front-end electronics and sending them to programmable peers on a commodity high bandwidth network. In the ATLAS Run 3 upgrade, FELIX will be used by the Liquid Argon (LAr) Calorimeters, Level-1 Calorimeter trigger system, BIS 7/8 and the New Small Wheel (NSW) muon detectors, as shown in the Fig.\,\ref{fig: felix_atlas}\cite{nsw_felix},\cite{lar_upgrade1}. In the ATLAS Run 4 upgrade, the FELIX approach will be used to interface with all ATLAS detector and trigger systems.

 FELIX brings multiple improvements in both performance and maintenance of the full DAQ (data acquisition) chain. Since the FELIX system maximizes the use of commodity hardware, the DAQ system can reduce its reliance on custom hardware. Furthermore additional COTS (commercial off-the-shelf) components can be easily connected to resize the FELIX infrastructure as needed. The FELIX system implements a switched network architecture which makes the DAQ system easier to maintain and more scalable for future upgrades\,\cite{felix_john}. The FELIX architecture meets the following requirements:

\begin{itemize}
	\item FELIX should be detector independent.
	\item FELIX must support the CERN standard GBT protocol with all its configuration options to connect to FE (Front-End) units having radiation hardness concerns\,\cite{gbt_manual}.
	\item FELIX must distribute TTC (Timing, Trigger and Control) signals via fixed latency optical links.
	\item FELIX must route data from different GBTx E-links to configurable network end-points.
	        E-links are low bandwidth (80 to 320\,Mb/s) serial electrical links that are aggregated into a single high speed (4.8\,Gb/s) GBT optical link.
	\item For the ATLAS Run 4 upgrade, FELIX should also support fast calibration operations for FE units, by implementing a mechanism to send control commands and distribute data packets simultaneously at high throughput, with a synchronisation mechanism that does not involve network traffic.
\end{itemize}

In this paper we introduce the FELIX hardware platform in Section\,\ref{sect: hardware}, the firmware design in Section\,\ref{sect: firmware} and software features in Section\,\ref{sect: software}. The status of integration activities with several ATLAS front-end units is described in Section\,\ref{sect: test}.

\section{The FELIX Interface Card}
\label{sect: hardware}
\begin{figure}[!t]
	\centering
	\includegraphics[width=3in]{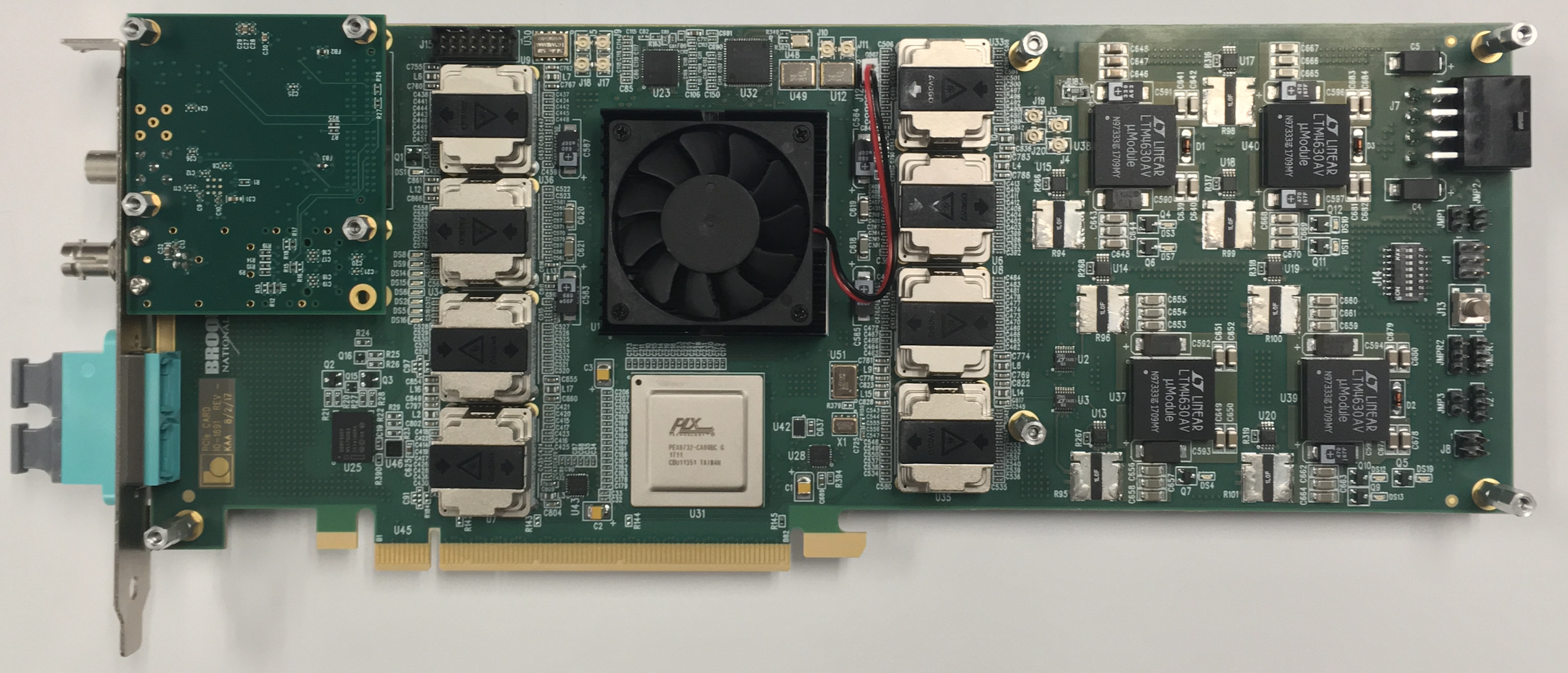}
	\caption{FELIX final prototype FPGA PCIe card (FLX-712)}
	\label{fig: flx_712}
\end{figure}

The FELIX hardware platform has been developed for the final implementation in the ATLAS Run 3 upgrade.  It is a standard height PCIe Gen3 card. The latest version is named as the FLX-712, as shown in Fig.\,\ref{fig: flx_712}. It is based on a Xilinx Kintex UltraScale FPGA (XCKU115-FLVF-1924) capable of supporting 48 bi-directional high-speed optical links via on-board MiniPOD transceivers, with a 16-lane PCIe Gen3 interface. In comparison to the previous version (FLX-711), the FLX-712 no longer hosts the unneeded DDR4 SODIMM connectors\,\cite{felix_kai}. This eases PCB routing and also makes the board shorter. Since the FPGA has two Super Logic Regions (SLRs), two 8-lane PCIe endpoints are implemented in separate SLRs to achieve a balanced placement and routing that allows more channels to be serviced and easier timing closure.

\begin{figure}[!h]
	\centering
	\includegraphics[width=0.95\columnwidth]{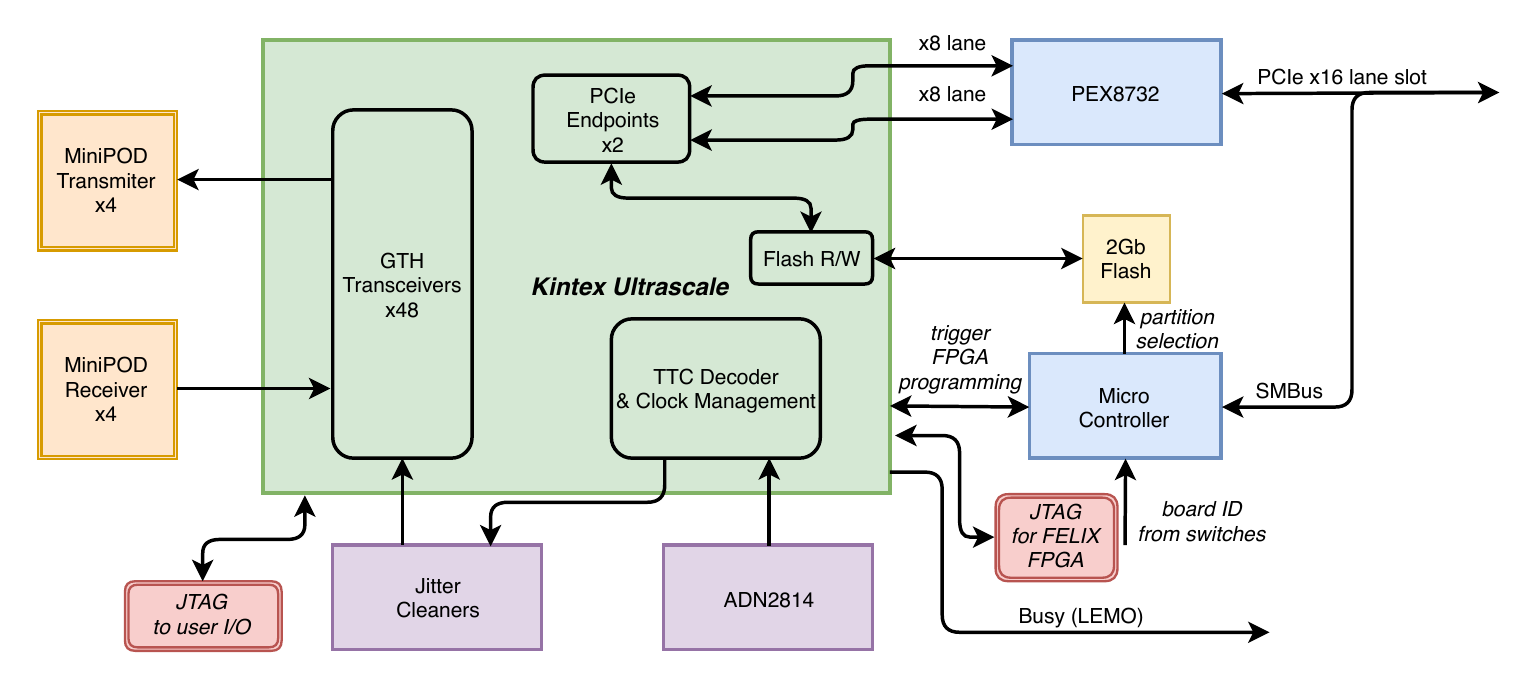}
	\caption{Block diagram of the FELIX final prototype FPGA card}
	\label{fig: flx_712_diagram}
\end{figure}

Fig.\,\ref{fig: flx_712_diagram} shows the functional block diagram of the FLX-712. Since the Xilinx UltraScale FPGA supports at most 8-lane PCI Express, a PCIe switch (PEX8732) is used to connect two 8-lane endpoints to the 16-lane PCIe slot. This approach ensures that it is possible to achieve the required nominal bandwidth of 128\,Gb/s. There are four transmitter MiniPODs and four receiver MiniPODs on board; each one has 12 high-speed Rx or Tx links connected to FPGA GTH transceivers\,\cite{ultrascale_gth}. The speed of these 48 optical links can be up to 14\,Gb/s, which is limited by the MiniPODs. An on-board jitter cleaner chip (Si5345) is used to provide a low jitter reference clock, at an integer multiple of the BC (bunch-crossing) clock, for the GTH transceivers. The Front-end optical links can connect to the FLX-712 via two optical multi-fiber (MTP) couplers. The MTPs can each be either MTP-24 (12 pairs) or MTP-48 (24 pairs) according to the application.

\begin{figure}[!h]
	\centering
	\includegraphics[width=3in]{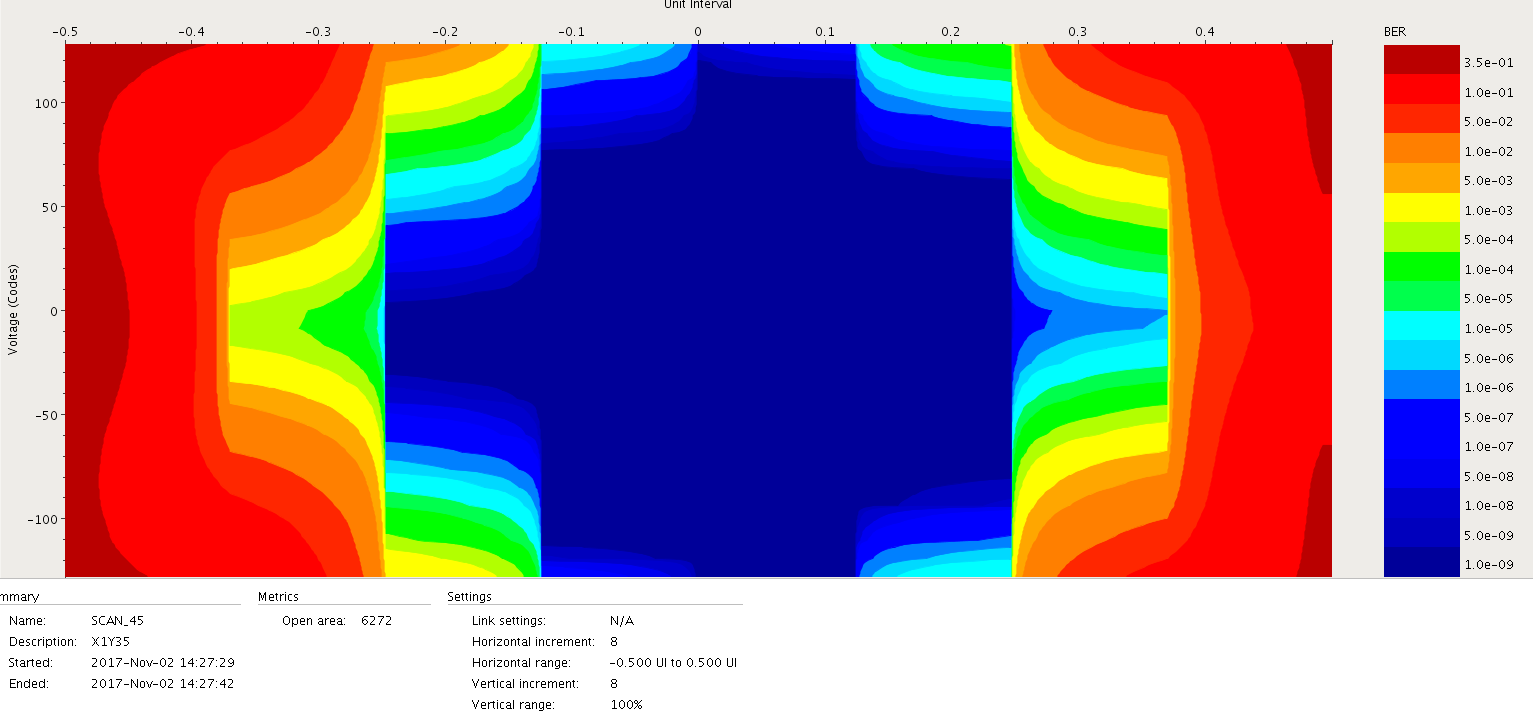}
	\caption{The eye diagram for one optical transceiver channel at 12.8\,Gb/s}
	\label{fig: flx_712_eye}
\end{figure}

 All of the hardware features of FLX-712 have been successfully verified. To test the PCIe interface, two Wupper DMA engines were implemented in the FPGA. Counter patterns were then used to test the throughput to the host server. The total measured throughput of these two 8-lane PCIe Gen3 endpoints can be up to 101.7\,Gb/s, in agreement with the PCIe specification. To test the optical links, the Xilinx IBERT IP was used to perform BER (Bit Error Rate) and eye diagram tests at line rates of 12.8\,Gb/s and 9.6\,Gb/s\,\cite{ultrascale_gth}\,\cite{ibert}. The results show that the BER is smaller than $10^{-15}$ for all of the 48 optical links. A typical eye diagram at 12.8\,Gb/s is shown in Fig.\,\ref{fig: flx_712_eye}.

\section{FELIX Firmware}
\label{sect: firmware}
\begin{figure}[!h]
	\centering
	\includegraphics[width=3in]{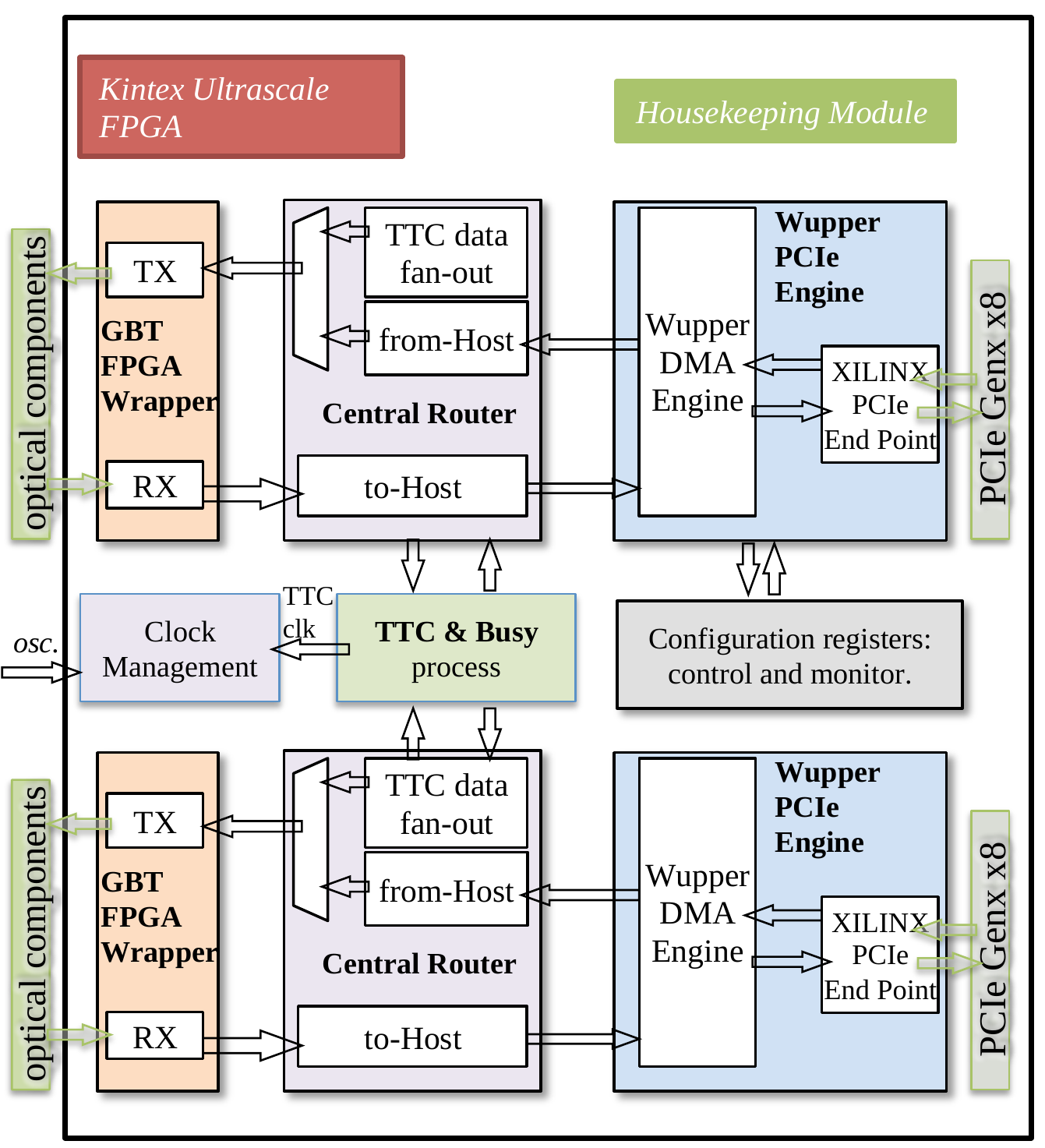}
	\caption{Block diagram of the FLX-712 firmware}
	\label{fig: fw_diagram}
\end{figure}

The FELIX firmware supports two modes: GBT mode and FULL mode. GBT mode uses GigaBit Transceiver (GBT) architecture and a protocol developed by CERN providing a bi-directional high-speed (4.8\,Gb/s) radiation-hard optical link\,\cite{gbt_manual}. FULL mode uses a customized light-weight protocol for the from front-end path, providing a higher maximum payload at a line rate of 9.6\,Gb/s. As FULL mode uses 8b/10b encoding, a maximum user payload of 7.68\,Gb/s can be achieved. The main functional blocks of the FELIX firmware, shown in Fig.\,\ref{fig: fw_diagram}, consist of a GBT wrapper, Central Router, PCIe Direct Memory Access (DMA) engine and other modules. Two sets of firmware modules are instantiated in the top level design to have a balanced structure and to ease FPGA net routing.

In addition to routing front-end data streams, FELIX also distributes TTC information to front-end electronics from the TTC system. The TTC decoder firmware module is based on the TTC firmware from the CERN GLIB project\,\cite{cern_glib}. It receives the clock and serial TTC data from a TTC optical fiber via a clock and data recovery chip (ADN2814). The serial TTC data contains two interleaved data streams: the A-channel, reserved for the Level-1 Accept, and the B-channel which carries other commands such as BCR (Bunch Counter Reset). The generated 40.08\,MHz TTC clock from the MMCM is distributed via a dedicated clock net to the rest of FPGA fabric. Due to the low jitter requirement of the high-speed GTH transceivers, their reference clock is provided by the on-board jitter cleaner (Si5345) which multiplies the frequency and cleans the jitter.

The FELIX GBT wrapper is based on the CERN GBT-FPGA firmware with several performance improvements\,\cite{gbt_fpga}. It encapsulates the Forward Error Correction (FEC) encoder/decoder, a scrambler/descrambler and a gearbox architecture. To decrease the latency, the frequency of the FEC encoder/decoder and scrambler/descrambler clock domain was increased to 240\,MHz\,\cite{gbt_fpga_optimization}. The GBT protocol supports GBT frame-encoding mode and wide-bus mode\,\cite{gbt_fpga}. The wide-bus mode is not radiation tolerant, as the FEC encoder and decoder are sacrificed in the to-host direction in favor of a higher user payload.

\begin{figure}[!h]
	\centering
	\includegraphics[width=0.95\columnwidth]{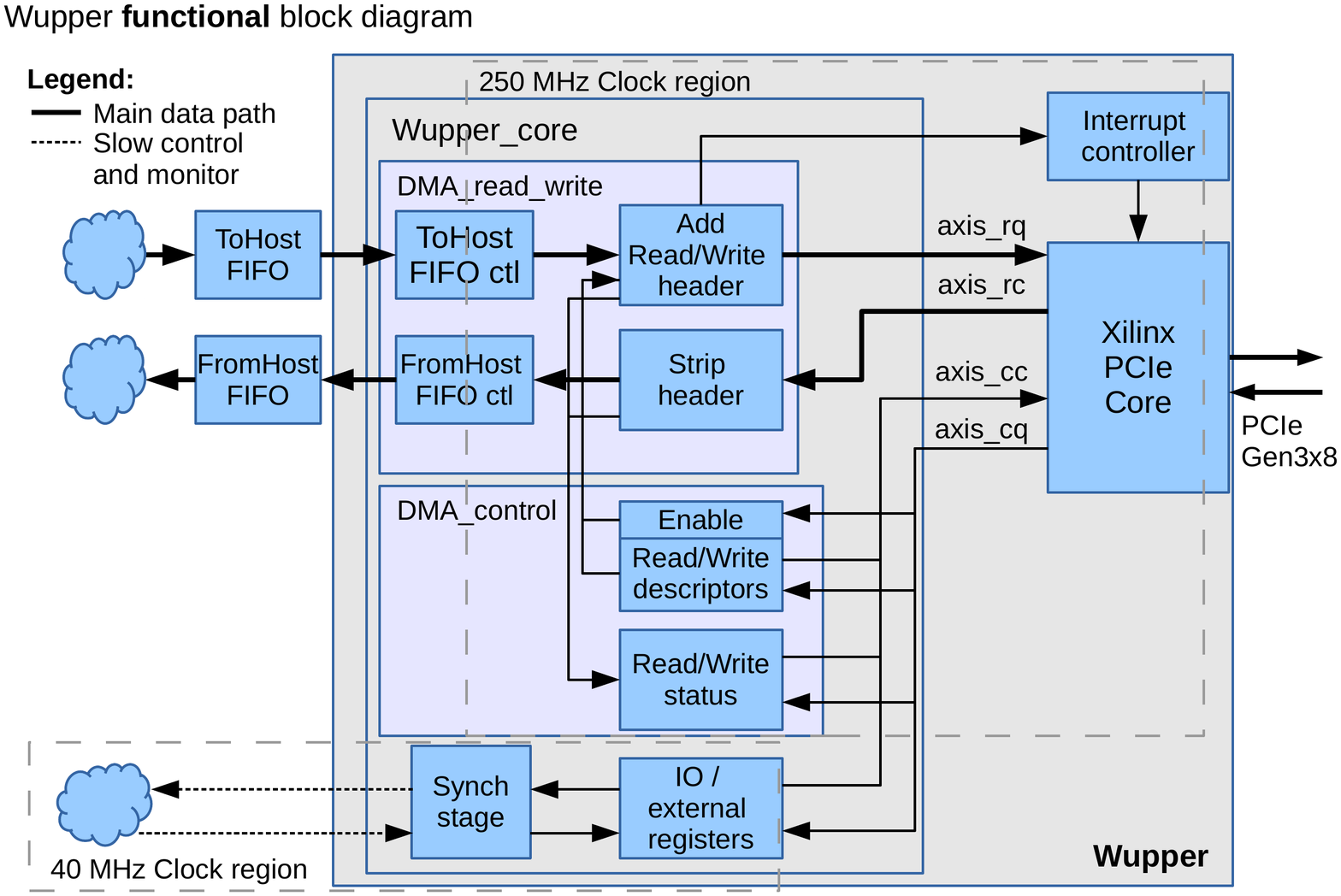}
	\caption{Structure of the Wupper PCIe engine}
	\label{fig: wupper}
\end{figure}

PCIe firmware, called Wupper, was designed to provide a simple Direct Memory Access (DMA) interface for the Xilinx PCIe Gen3 hard block\cite{felix_pcie}. It transfers data between a 256-bit wide user logic FIFO and the host server memory, according to the addresses specified in DMA descriptors. Up to eight descriptors can be queued to be processed sequentially. Since the Xilinx PCIe Gen3 hard block only supports a maximum of eight lanes, the FPGA implements two 8-lane PCIe endpoints with separate DMA engines. The block diagram of the Wupper design is shown in Fig.\,\ref{fig: wupper}. Its functional blocks can be categorized into two groups: DMA control and DMA write/read. The DMA control parses and monitors received descriptors. The DMA write/read blocks process the data streams for both directions. If the received descriptor is a to-host descriptor, the payload data is read from the user logic FIFO and added after the header information. If the descriptor is a from-host descriptor, the header of received data is removed and the length is checked; then the payload is shifted into the FIFO.

\section{FELIX Software}
\label{sect: software}
The FELIX software suite has different layers: for example, low-level software tools, test software and production software. Access to the FELIX hardware level is controlled via two device drivers: flx and cmem\_rcc. The flx driver is a conventional character driver for PCIe interface cards. Its main function is to provide virtual addresses for the registers of a FLX-712 card that can be used directly by user processes for access to the hardware. This design avoids the overhead of a context switch per IO transaction and is therefore essential for the performance of FELIX. The cmem\_rcc driver, from the ATLAS TDAQ project, allows the application software to allocate large buffers of contiguous memory. For use with FELIX, it has been tested for buffers of up to 16\,GByte and the allocation time of large buffers has been reduced.

\begin{figure}[!h]
	\centering
	\includegraphics[width=3in]{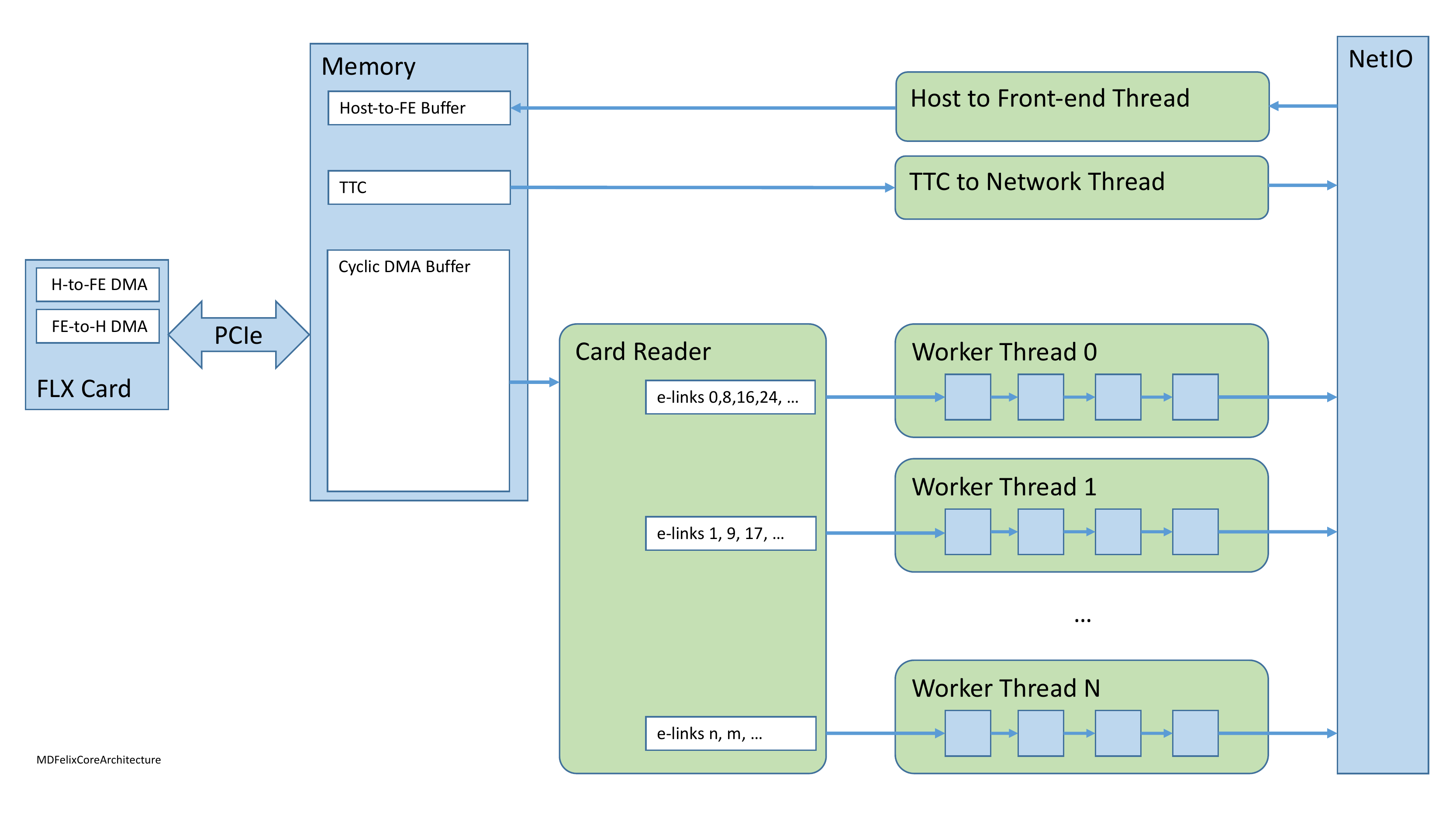}
	\caption{The felixcore application architecture}
	\label{fig: core_application}
\end{figure}

The felixcore application handles the data between the front-ends using the FLX-712 card and a dedicated library called NetIO. Its functional architecture is shown in Fig.\,\ref{fig: core_application}. It does not perform any content analysis or manipulation of the data, other than that which is needed for decoding and transport. The DMA engine transfers a data stream into a contiguous circular buffer which is allocated using the cmem\_rcc driver in the memory of the host server. Continuous DMA enables data transfer at full speed and does not require the DMA to be re-set for each transfer. Data blocks retrieved from the circular buffer are inspected for integrity while extracting the E-link identifier and sequence number. The block is then copied to a selected worker thread based on the E-link identifier. The worker threads recombine the data stream for each E-link if any splitting for transport proved necessary. Once the data reconstruction is complete, the data are appended with a FELIX header and published to the network through NetIO.

NetIO is implemented as a generic message-based networking library that is tuned for typical use cases in DAQ systems. It offers four different communication modes: low-latency point-to-point communication, high-throughput point-to-point communication, low-latency publish/subscribe communication and high-throughput publish/subscribe communication. NetIO has a backend system to support different network technologies and API's. At this  time, two different backends exist. The first backend uses POSIX sockets to establish reliable connections to endpoints. Typically this backend is used for TCP/IP connections in Ethernet networks. The second backend uses libfabric for communication and is used for Infiniband and similar network technologies\,\cite{libfabric}. Libfabric is a network API that is provided by the OpenFabrics Working Group. There are six different user-level sockets in NetIO, of which four are point-to-point sockets (one send socket and one receive socket, each in a high-throughput and a low-latency version), and two publish/subscribe sockets (one publish and one subscribe socket). The publish/subscribe sockets internally use the point-to-point sockets for data communication.

\begin{figure}[!h]
 	\centering
 	\includegraphics[width=3in]{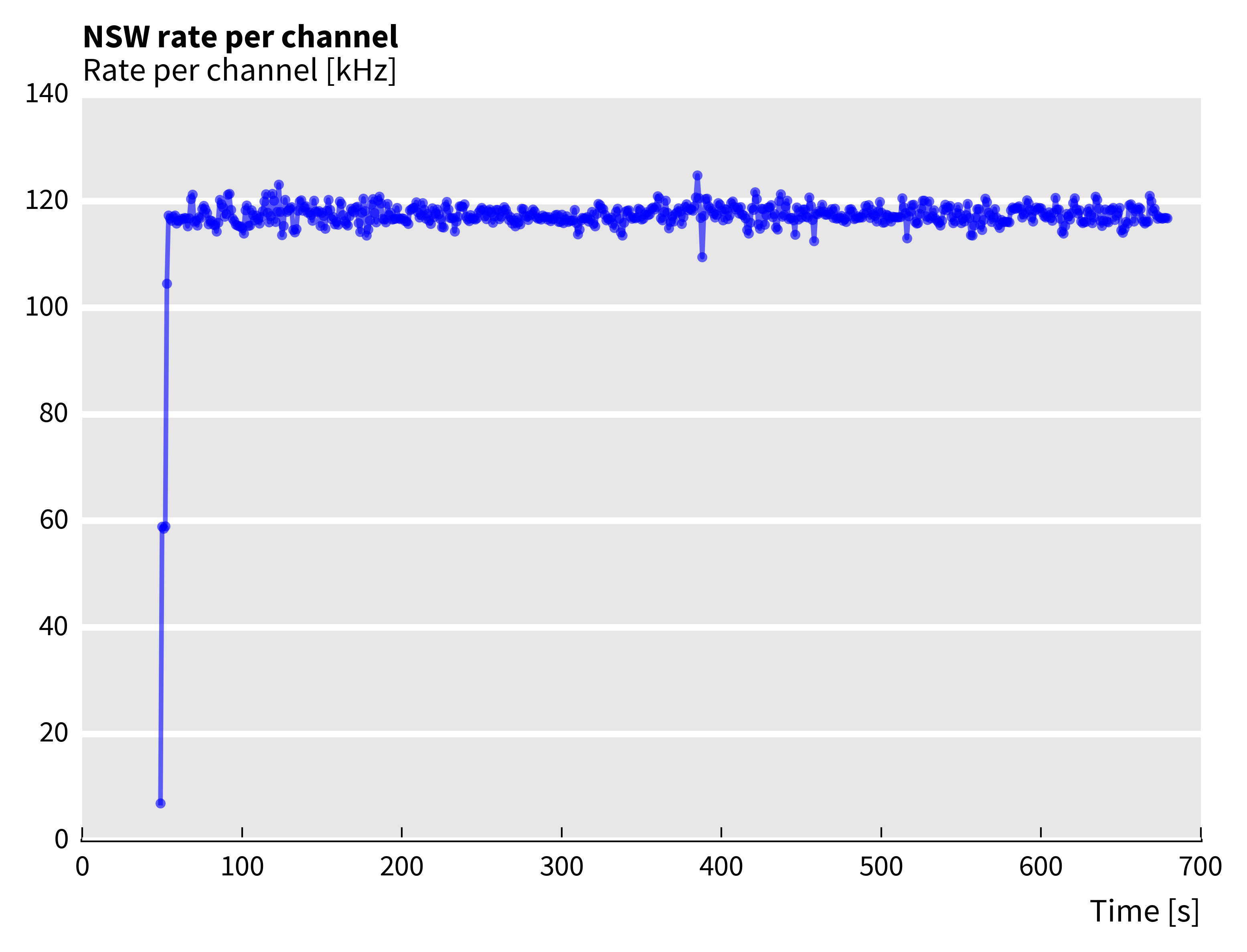}
 	\caption{Felixcore performance for GBT mode}
 	\label{fig: gbt_performance}
 \end{figure}

A number of benchmarks have been carried out to evaluate the performance of felixcore application and NetIO. These tests were run with a host server as the FELIX and another host as the data receiver. A 40 GbE connection was available between the hosts. 
In the GBT mode performance test, two FLX-712 cards were used to support 48 GBT links. The FLX cards were configured to the most demanding workload for the ATLAS Run 3 upgrade, with 8 E-links per GBT link and a chunk size of 40 Bytes. As shown in Fig.\,\ref{fig: gbt_performance}, the system is comfortably able to transfer the full load at above the ATLAS L1 Accept rate of 100 kHz. Benchmarking for the FULL mode case also indicates that it will be possible to handle data at the L1 Accept rate.

\begin{figure}[!h]
	\centering
	\includegraphics[width=3in]{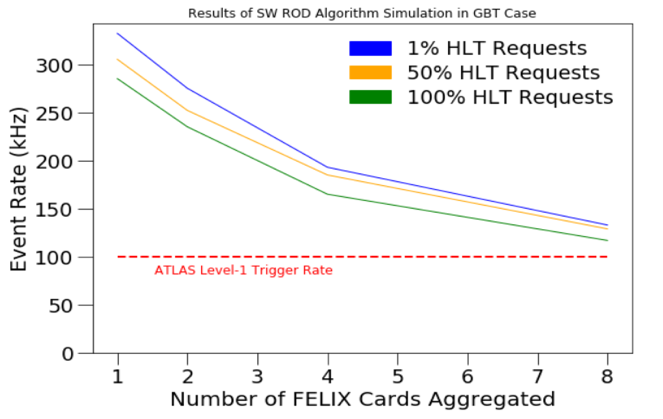}
	\caption{Software ROD performance}
	\label{fig: sw_rod}
\end{figure}

A Software ROD (ReadOut Driver) is an application running on a commodity server which receives data from one or more FELIX systems and performs flexible data aggregation and formatting tasks. Incoming data packets associated with a given ATLAS event are automatically logically aggregated into a larger event fragment for further processing.
The data are finally formatted to match common ATLAS specification, as produced by existing readout system, for consumption by High Level Trigger (HLT) on request. Benchmarks for the current aggregation algorithms, including realistic simulation of the cost of subdetector processing and HLT request handling, were carried out with simulated input data from multiple FELIX cards, each with 192 E-links and realistic packet sizes. The test results are shown in the Fig.\,\ref{fig: sw_rod}. The algorithm is shown to be able to handle input from multiple FELIX cards, with the performance able dependent on host CPU speed and number of cores. The 1\%, 50\% and 100\% in the plot refer to the fraction of the events arriving at the software ROD which the HLT then samples. 

\section{Integration tests with different front-ends}
\label{sect: test}
For the upcoming ATLAS Run 3 upgrade in 2019, FELIX will be implemented to interface with several detector front-ends, such as the Muon Spectrometer's New Small Wheel (NSW), Liquid Argon Calorimenter (LAr) Trigger Digitizer Board (LDPB) and the Level-1 Calorimeter Trigger (L1Calo) system\,\cite{nsw_felix}\,\cite{gfex_hucheng}. For the Run 4 upgrade of HL-LHC (High-Luminosity LHC), the plan is to adopt FELIX to interface with all the detector front-ends.

\subsection{Integration Test with New Small Wheel Front-ends}
In the NSW integration tests, FELIX successfully distributed TTC information to front-end electronics, including the bunch crossing clock and L1A trigger signal. The dataflow to and from the front-ends has been demonstrated. FELIX can also trigger a front-end test pulse from a test application, and successfully configure ASICs and FPGAs via the GBT-SCA's GPIO, \ItwoC, SPI and JTAG interfaces\, \cite{gbt_sca}. Other highlights also include the ability to read out ADC monitoring data and configure the GBTx on the L1DDC board\,\cite{l1ddc}. Taken together these tests provide a robust demonstration of the functionality of the IC and SCA links in the GBT frame\,\cite{gbt_manual}.

\subsection{Integration Test with Liquid Argon Calorimeter LTDB}
\begin{figure}[!h]
	\centering
	\includegraphics[width=3in]{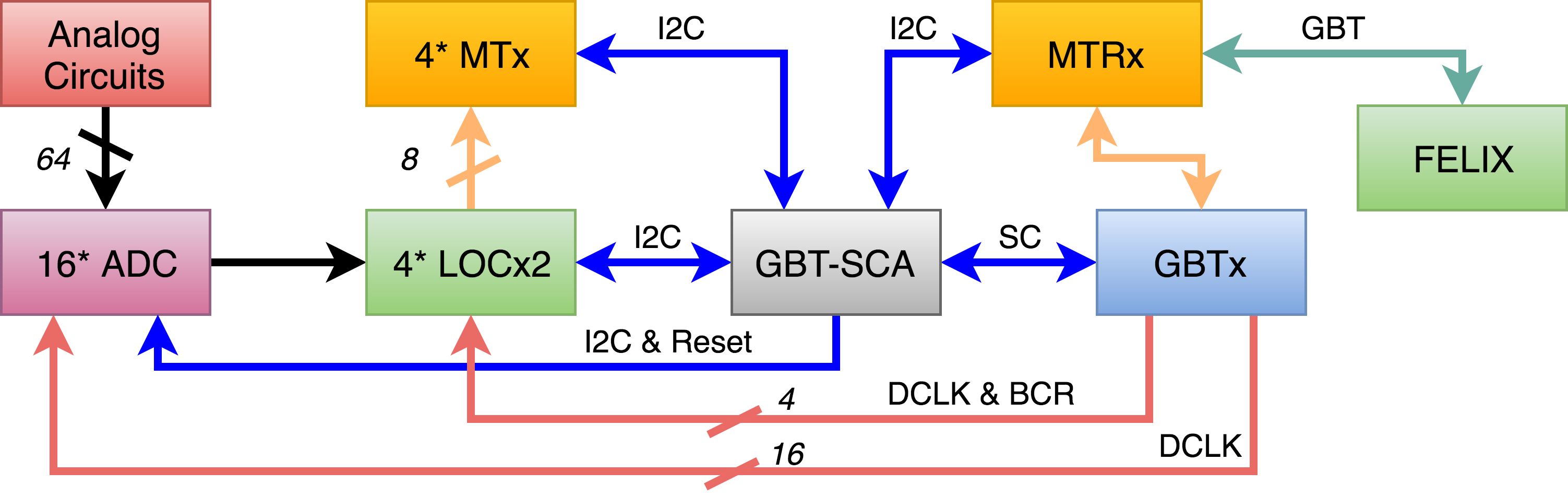}
	\caption{GBTx and GBT-SCA connections in the LTDB prototype}
	\label{fig: ltdb}
\end{figure}

In the LAr (Liquid Argon Calorimeter) Run 3 upgrade, the LAr Trigger Digitizer Board (LTDB) digitizes input analog signals, and transmits them to the back-end system\,\cite{lar_upgrade1}. There are five GBTx and five GBT-SCA chips on the LTDB prototype. And five GBT links in total from FELIX are connected to the LTDB. Part of the connection scheme (one GBT link) is shown in Fig.\,\ref{fig: ltdb}. GBT-SCA chips are used to control the power rails, \ItwoC buses and also perform on-board temperature measurement\,\cite{gbt_sca}. Besides the interface to EC links with the GBT-SCA chip, each GBTx on the LTDB provides the recovered 40\,MHz TTC clock from a FELIX GBT link to the ASICs of the NEVIS ADC and serializers LOCx2, and also sends the BCR (Bunch Counter Reset) signal to the LOCx2 ASIC\,\cite{nevis_adc}\,\cite{locx2}. 

\subsection{Integration Test with gFEX}
\begin{figure}[!h]
	\centering
	\includegraphics[width=3in]{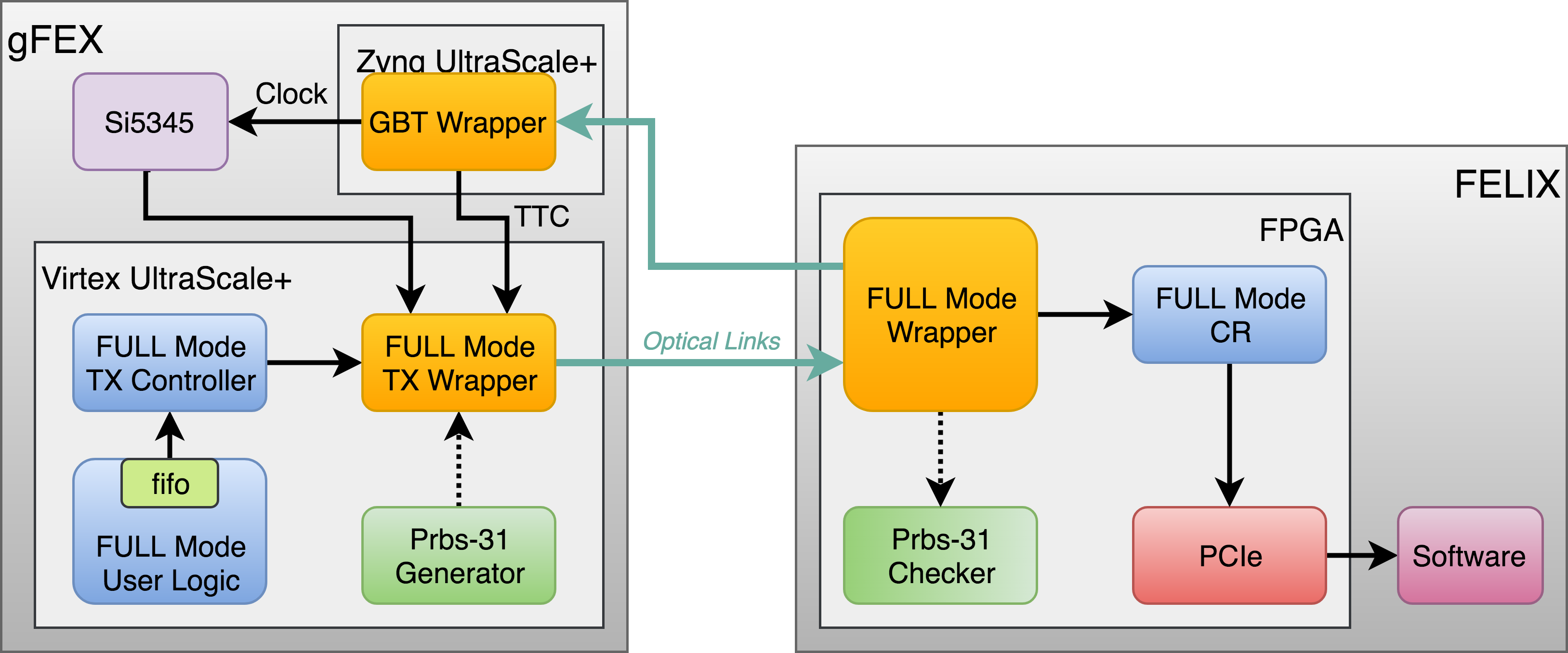}
	\caption{Block diagram of gFEX and FELIX integration test}
	\label{fig: gfex_felix}
\end{figure}

The Global Feature Extractor (gFEX) is one of several modules that will be deployed in the Level-1 Calorimeter (L1Calo) trigger system in the ATLAS Run 3 upgrade\,\cite{gfex_tang}. In the integration test of gFEX and FELIX, gFEX needs to recover the TTC clock from a FELIX GBT link at 4.8\,Gb/s, and also receive TTC signals such as Level-1 trigger accept and BCR. As for the to-host path, gFEX needs to send data to FELIX using FULL mode optical links at 9.6\,Gb/s. A block diagram of the test setup is shown in Fig.\,\ref{fig: gfex_felix}. The test results show that gFEX recovers a stable TTC clock and receives the TTC information correctly. The latency of TTC signal transmission (from TTC system to gFEX through FELIX) is fixed and does not change under conditions such as transceiver reset, fiber reconnection, TTC system power cycling and FELIX \& gFEX power cycling. The FULL mode links from gFEX to FELIX have been tested with the PRBS-31 (Pseudo Random Bit Sequence) data pattern. No error was observed and the BER (Bit Error Rate) is smaller than $10^{-15}$.

\section{Conclusion}
FELIX is a readout system that interfaces custom links from front-end electronics to standard commercial networks in the ATLAS upgrade. FELIX also distributes the LHC bunch-crossing clock, trigger accepts and resets received from the TTC system to detector front-ends through fixed latency optical links. It supports the CERN standard 4.8\,Gb/s GBT protocol and a customized lightweight FULL mode which has a higher throughput of 9.6\,Gb/s. The results of integration and performance tests with ATLAS front-end systems to date indicate that FELIX is on course to be ready for deployment in 2019.

%\appendices
%\section{Proof of the First Zonklar Equation}
%Appendix one text goes here.
%% you can choose not to have a title for an appendix
%% if you want by leaving the argument blank
%\section{}
%Appendix two text goes here.

% use section* for acknowledgment
%\section*{Acknowledgment}
%The authors would like to thank...

% Can use something like this to put references on a page
% by themselves when using endfloat and the captionsoff option.
\ifCLASSOPTIONcaptionsoff
  \newpage
\fi

%\bibliographystyle{plain}
%\bibliography{wwbib}

\end{document}